\documentclass[%
twocolumn,
 amsmath,amssymb,
 aps,
pra,
]{revtex4-1}

\usepackage{graphicx}
\usepackage{dcolumn}
\usepackage{bm}
\usepackage{subfigure}
\usepackage[mathlines]{lineno}
\usepackage{natbib}
\usepackage{braket}
\usepackage{multirow}
\usepackage{bbold}

\begin{document}


\title{Collisional EPR Frequency Shifts in Cs-Rb-Xe Mixtures}

\author{S. Zou}
\altaffiliation[Present address: ]{Beihang University, 37 Xueyuan Road, Haidian District, Beijing, 100191, China}

\author{D.J. Morin}
\affiliation{Department of Physics and Astronomy, Washington State University, P.O. Box 642814, Pullman, WA 99164-2814, USA}

\author{C. Weaver}
\affiliation{Department of Physics and Astronomy, Washington State University, P.O. Box 642814, Pullman, WA 99164-2814, USA}


\author{Z. Armanfard}
\affiliation{Department of Physics and Astronomy, Washington State University, P.O. Box 642814, Pullman, WA 99164-2814, USA}


\author{J. Muschell}
\affiliation{Department of Physics and Astronomy, Washington State University, P.O. Box 642814, Pullman, WA 99164-2814, USA}

\author{A.I. Nahlawi}
\affiliation{Department of Physics and Astronomy, Washington State University, P.O. Box 642814, Pullman, WA 99164-2814, USA}

\author{B. Saam}
\email{brian.saam@wsu.edu}
\affiliation{Department of Physics and Astronomy, Washington State University, P.O. Box 642814, Pullman, WA 99164-2814, USA}

\date{\today}

\begin{abstract}

We report a measurement of the ratio of dimensionless enhancement factors $\kappa_0$ for Cs-$^{129}$Xe and Rb-$^{129}$Xe in the temperature range $115-140$~$^{\circ}$C; both pairs are used in spin-exchange optical pumping (SEOP) to produce hyperpolarized $^{129}$Xe. $\kappa_0$ characterizes the amplification of the $^{129}$Xe magnetization contribution to the alkali-metal electronic effective field, compared to the case of a uniform continuous medium in classical magnetostatics. The measurement was carried out in ``hybrid" vapor cells containing both Rb and Cs metal in a prescribed ratio, producing approximately the same vapor density for both. Alternating measurements of the optically detected electron-paramagnetic-resonance (EPR) frequency shifts caused by the SEOP polarization and subsequent sudden destruction of the same quantity of $^{129}$Xe magnetization were made for $^{133}$Cs and either $^{87}$Rb or $^{85}$Rb. An important source of systematic error caused by power fluctuations in the pump laser that produced variable light shifts in the EPR frequency was characterized and then mitigated by allowing sufficient warm-up time for the pump laser. We measured $(\kappa_0)_{\rm CsXe}/(\kappa_0)_{\rm RbXe} = 1.215 \pm 0.007$ with no apparent temperature dependence. Based on our previous measurement $(\kappa_0)_{\rm RbXe}=518 \pm 8$, we determine $(\kappa_0)_{\rm CsXe} = 629 \pm 10$, more precise than but consistent with a previous measurement made by another research group.
\end{abstract}
\maketitle

\section{\label{sec:intro}Introduction}
Noble gases having non-zero nuclear spin, especially the stable spin-1/2 isotopes $^{3}{\rm{He}}$ and $^{129}{\rm{Xe}}$, are readily hyperpolarized to levels exceeding 10$\%$ via spin-exchange optical pumping (SEOP) \cite{Walker1997}, making them accessible to a wide variety of magnetic resonance experiments and applications \cite{Gentile2017,Meersman2014XeBook}. Isotopes with spin greater than 1/2, e.g., $^{83}$Kr and $^{131}$Xe, can also be hyperpolarized, albeit typically to lower levels due to faster nuclear relaxation \cite{Meersman2014XeBook,Stupic2011,Pavlovskaya2005}. The Fermi-contact hyperfine interaction between the alkali-metal electron and the noble-gas nucleus is crucial to SEOP physics but still incompletely understood, especially for the heavier noble gases. The heavier noble gases have advantages in searches for physics beyond the Standard Model: searches for a permanent electric-dipole-moment (EDM) \cite{Allmendinger2019,Sachdeva2019}, spin-mass interactions \cite{Almasi2018}, and new axion-like particles \cite{Bulatowicz2013} have all made use of optically pumped and polarized $^{129}$Xe and/or $^{131}$Xe. There is renewed interest in hyperpolarized $^{131}$Xe to study time-reversal violation in neutron-$^{131}$Xe interactions \cite{Molway2021}. SEOP is also being extended to the even heavier but radioactive radon isotopes for greater EDM sensitivity \cite{Tardiff2014,Kitano1988}.

Efficient production of liter quantities of hyperpolarized xenon \cite{Driehuys1996,Ruset2006,Nikolau2013} is especially important for magnetic resonance imaging applications \cite{Roos2015,Ruppert2019,Grist2021}. One unresolved question in this context is whether Rb, Cs, or a mixture of the two (in a so-called ``hybrid" cell) is the ideal alkali-metal partner for Xe. While some interesting and suggestive results have been reported \cite{Whiting2011}, a complete answer to this question requires careful measurement of fundamental physical parameters such as spin-exchange efficiency for both SEOP pairs. Optically detected EPR of the alkali-metal hyperfine structure is a powerful and sensitive technique for characterizing a large fraction of basic SEOP physics: EPR frequency shifts can be used to monitor SEOP transients and measure spin-exchange rates. In a pulsed-EPR implementation that we are currently developing, both ``relaxation in the dark" \cite{Franzen1959,Nelson2001} and characterization of the alkali-metal hyperfine linewidth \cite{Appelt1999} as a function of buffer-gas pressure can yield accurate spin-destruction rates. In the limit where the hyperfine populations are described by a spin-temperature distribution \cite{Appelt1998} (which holds across a broad set of commonly realized experimental conditions), acquisition of the hyperfine spectrum allows a quantitative measurement of the alkali-metal polarization. Finally, absolute EPR shifts can be used to determine the noble-gas polarization, provided one has independently measured $\kappa_0$, a dimensionless enhancement factor that characterizes the overlap of the alkali-metal-electron wave function with the noble-gas nucleus during spin-exchange collisions. Measurements of the noble-gas polarization are, in turn, important for determining spin-exchange rate coefficients; our research group is interested, for example, in determining whether Rb or Cs is a more efficient SEOP partner for $^{129}$Xe. Knowledge of $\kappa_0$ is also important for many of the precision measurements discussed above, since the associated frequency shifts represent systematic effects that must be measured or eliminated \cite{Korver2015}.

We have previously measured $\kappa_0$ for the Rb-Xe pair by comparing both nuclear-magnetic-resonance (NMR) \cite{Ma2011} and EPR \cite{Nahlawi2019} frequency shifts in Rb-vapor cells containing both $^3$He and $^{129}$Xe, where $\kappa_0$ for Rb-$^3$He is well understood from a previous measurement \cite{Romalis1998}. In the present work, the only polarizable noble gas is $^{129}$Xe, but the cells contain both Rb and Cs. By acquiring alternating sequential measurements of the Rb and Cs EPR frequency shifts generated by the destruction of the same quantity of $^{129}$Xe magnetization (reproducibly generated by SEOP), we have made a precise measurement of the ratio $(\kappa_0)_{\rm CsXe}/(\kappa_0)_{\rm RbXe}$.

\section{\label{sec:theory}Theoretical Background}

We summarize here a more detailed theoretical discussion found in Refs.~\cite{Nahlawi2019} and \cite{Schaefer1989}. The Fermi-contact interaction essential for SEOP generates a shift $\Delta f_{\rm A}$ in the alkali-metal EPR frequency that is proportional to the noble-gas nuclear magnetization:

\begin{equation}
\label{eq:EPRshift}
\Delta|f_A|=\Bigg|\frac{df_A}{dB_0}\Bigg|\Bigg(\frac{8\pi}{3}\kappa_{AX}\mu_{\rm X}\frac{\langle K_z\rangle}{K}[{\rm X}]\Bigg),
\end{equation}

\noindent where $[{\rm X}]$, $\mu_{\rm X}$, and $K$ are the noble-gas number density, magnetic moment, and spin, respectively; $B_0$ is the applied magnetic field, and $\kappa_{\rm AX} > 0$ is a dimensionless factor specific to each alkali-metal/noble-gas pair that parameterizes the enhancement of the noble-gas magnetic field sensed by the alkali-metal electron. The frequency shift in Eq.~(\ref{eq:EPRshift}) can be understood as the product of the time-averaged magnetic field (in large parentheses) generated over many collisions by the noble-gas nuclei and the gyromagnetic ratio

\begin{equation}
\label{eq:ratio2}
\gamma_{\rm A}(F,\overline{m}_F, B_0)\equiv 2\pi\frac{df_{\rm A}}{dB_0},
\end{equation}

\noindent where we note that $\gamma_{\rm A}$ is only valid locally for small field changes around a specific value of $B_0$ for a specific hyperfine transition. Here, we have defined the total ground-state alkali-metal spin as ${\mathbf F}={\mathbf S}+ {\mathbf I}$, the sum of the electron and nuclear spins, respectively, and $\overline{m}_F$ as the mean azimuthal quantum number for the neighboring levels involved in the transition.

If the respective alkali-metal EPR frequency shifts for Cs and Rb due to the complete destruction of the same quantity of Xe magnetization are measured, then we can form the ratio:

\begin{equation}
\label{eq:ratio1}
\frac{\Delta f_{\rm Cs}}{\Delta f_{\rm Rb}} = \Bigg(\frac{\gamma_{\rm Cs}}{\gamma_{\rm Rb}}\Bigg) \frac{(\kappa_0)_{\rm CsXe}}{(\kappa_0)_{\rm RbXe}}.
\end{equation}

\noindent Equation (\ref{eq:ratio1}) applies equally well but separately to the two isotopes $^{85}$Rb and $^{87}$Rb. Gas densities in our vapor cells were sufficiently high that the limit $\kappa_{\rm AX}=\kappa_{\rm XA} \rightarrow \kappa_0$ was valid \cite{Nahlawi2019,Schaefer1989}, where $\kappa_{\rm AX}$ and $\kappa_{\rm XA}$ are the enhancement factors for the EPR and NMR shifts, respectively. The goal of this work is to measure the ratio on the left-hand side of Eq.~(\ref{eq:ratio1}) in order to extract the $\kappa_0$-ratio on the right-hand side. For our experiments $B_0$ is large enough that the quadratic Zeeman term in the Breit-Rabi equation \cite{Breit1931} resolves the hyperfine transition lines. Under optical pumping conditions, this spectrum is most intense at the end resonance $\overline{m}_F=\pm I$ of the $F=I+1/2$ manifold corresponding to $\sigma\pm$ pumping light. Our experiments utilize these two resonance lines, for which \cite{Breit1931}

\begin{equation}
\label{eq:gammap}
\gamma_A\Big(I+\frac{1}{2}, \pm I, B_0\Big)=\gamma_0\Bigg[1+2I\frac{g_I}{g_S}\pm 4I\frac{\gamma_0B_0}{2\pi \delta_A}\Bigg],
\end{equation}

\noindent correct to terms linear in $B_0$. In the limit $B_0\rightarrow 0$ and for $|g_I|\ll |g_S|$, the field-independent gyromagnetic ratio for all nearest-neighbor transitions is

\begin{equation}
\label{eq:gamma0}
\gamma_0= \frac{g_S\mu_B}{2I+1}.
\end{equation}

\noindent In Eqs.~(\ref{eq:gammap}) and (\ref{eq:gamma0}), $\mu_B$ is the Bohr magneton, $g_S = -2.0023$ is the electron g-factor, $g_I$ is the nuclear g-factor (referenced to $\mu_B$), and $\delta_A$ is the precisely known alkali-metal hyperfine splitting \cite{SteckRb87-2015,SteckRb85-2013,SteckCs133-2010}.

\section{\label{sec:expt}Experiment}

\begin{figure}
\includegraphics[width=\columnwidth]{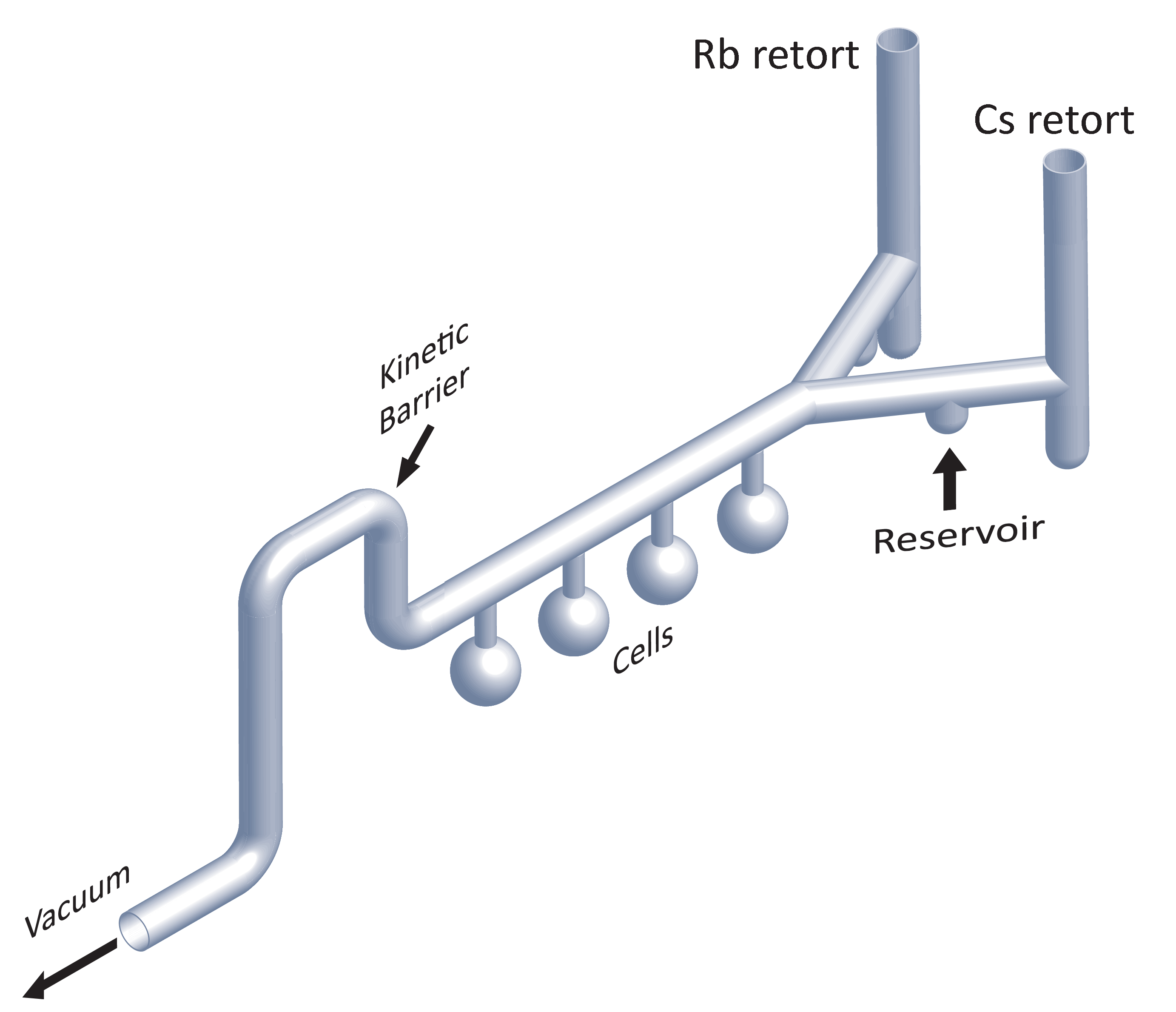}
\caption{\label{fig:glass} Glass manifold used to make Rb/Cs hybrid vapor cells. The two retorts are used to distill in Cs and Rb metals separately into each cell. The reservoirs capture the alkali metal distilled from the retorts prior to sealing the retort away from the manifold. The cells are then each filled with the prescribed gas mixture before being sealed away from the manifold while submerged in liquid nitrogen (to allow final room-temperature pressures $> 1$~atm). The kinetic barrier prevents alkali-metal from streaming directly into the vacuum system.}
\end{figure}

In order to probe the $^{129}$Xe magnetization with both Rb (either isotope) and Cs EPR, we fabricated so-called ``hybrid" vapor cells, which contained macroscopic amounts of Cs and Rb metals in a prescribed ratio targeted to generate approximately equal vapor densities of $10^{12}-10^{13}$~cm$^{-3}$ at $100-120$~$^{\circ}$C. The cells were fabricated and filled with an in-house gas-handling and vacuum system \cite{Jacob2002}; the cell properties are listed in Table~\ref{tab1}. The desired volume ratio of Cs to Rb metals ($\approx 8$ in our case) is estimated from known vapor-density curves \cite{Killian1926,CRC} and Raoult's Law \cite{Raoult1886}; see also the Supplemental Material \cite{SM}. A diagram of the manifold with attached cells fabricated by our glass blower is shown in Fig.\ \ref{fig:glass}. Whereas our typical manifold would have a single retort at the upstream end for a single alkali metal, this manifold was branched to include two retorts. With the manifold evacuated we used a flame to chase a small quantity ($\sim 1$~mg) of Cs metal into all of the cells, allowed everything to cool, and then chased Rb metal in to each cell prior to adding the requisite gases and flame-sealing each cell from the manifold. The volume of each metal in all cells was estimated ($\pm 15\%$) by noting the shape of the melt, measuring its dimensions, and applying an approximate geometric model. Care was taken to keep the metals separate within the cell until after the respective volume estimates were made.

\begin{table}
\caption{\label{tab1}
Gas composition (in torr at 20~$^{\circ}$C) of the three cells used in this work. Cell volume in cm$^3$ is shown in parenthesis below the cell designation. The estimated volume ratio $V_{\rm Rb}/V_{\rm Cs}$ is given in the top row, followed by the constituent gas and then total gas densities. The xenon (Spectra Gases) is enriched to 90\% $^{129}$Xe; the fraction of $^{131}$Xe is $0.2\%$.}
\begin{ruledtabular}
\begin{tabular}{lllllllll }
&{Cell 304A}&{Cell 304B}&{Cell 304C}\\
&{($4.24\pm 0.10$)}&{($3.99\pm 0.10$)}&{($4.12\pm 0.07$)}\\
\\[-0.9em]
\hline
\\[-0.6em]
$V_{\rm Rb}/V_{\rm Cs}$    &  8.3 $\pm$ 2.9   &  8.3 $\pm$ 3.7   &   10.0 $\pm$ 3.7\\
Xe    & 40.1 $\pm$ 1.3 		& 40.4 $\pm$ 1.4 		& 32.5 $\pm$ 1.4	\\
${^4}$He  & 1912 $\pm$ 52	& 1932 $\pm$ 53		& 1951 $\pm$ 53		\\
N$_2$& 57.7 $\pm$ 1.6		& 57.0 $\pm$ 1.6		& 56.6 $\pm$ 1.5	\\
Total (gas)  & 2010 $\pm$ 55	& 2029 $\pm$ 56		& 1986 $\pm$ 53		\\
\end{tabular}
\end{ruledtabular}
\end{table}


\subsection{\label{sec:odepr}Optically Detected EPR}

\begin{figure*}
\includegraphics[width=\textwidth]{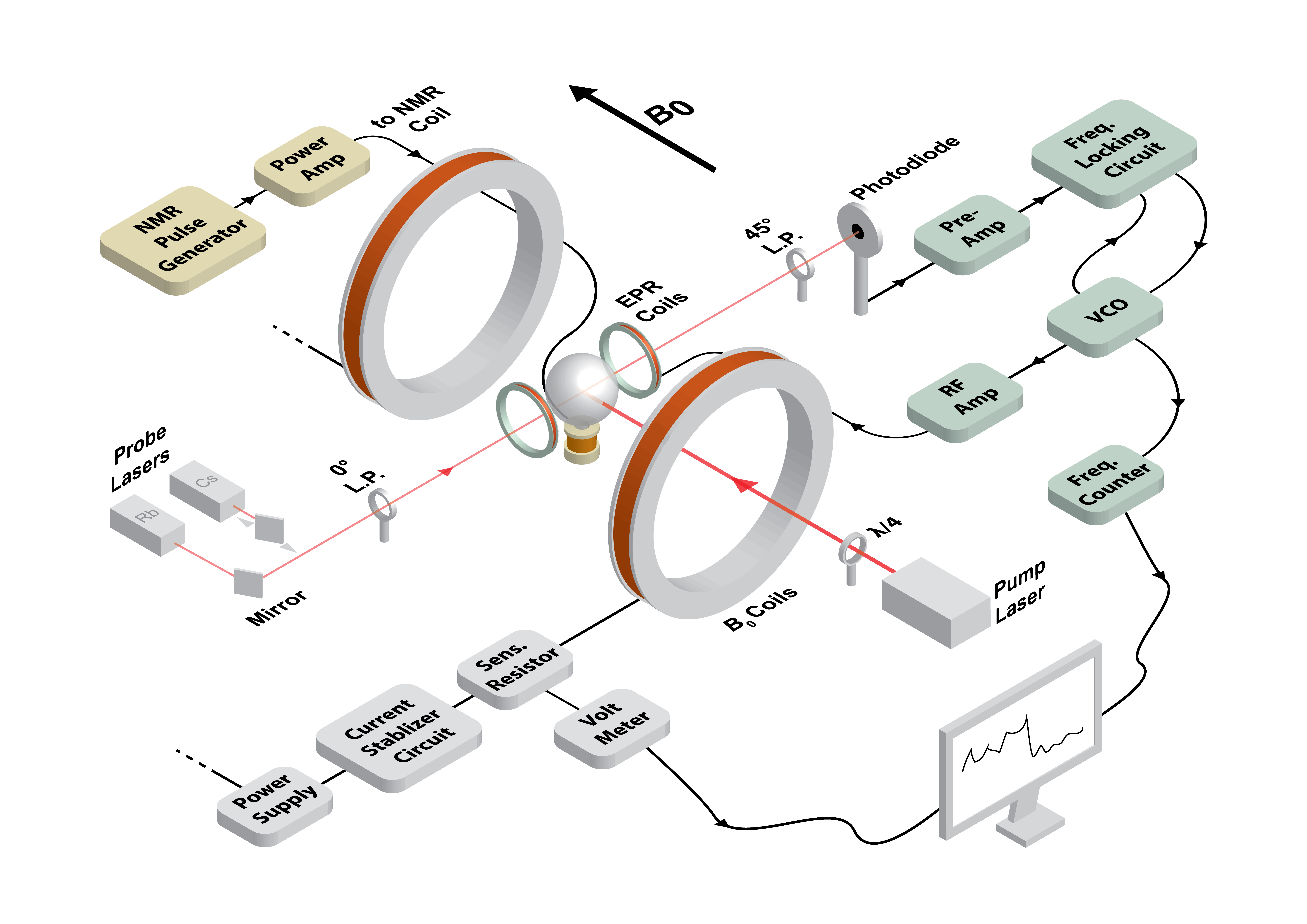}
\caption{\label{fig:app} Schematic diagram of experimental apparatus. More detail for the locking circuitry is shown in Ref.\ \cite{Nahlawi2019}. The main changes from that work are the orientation and location of the NMR coil, which was used only for transmission of pulses to destroy $^{129}$Xe magnetization, and the use of dual probe lasers (one each for Rb and Cs EPR) with a pick-off mirror positioned so that the lasers could be reproducibly swapped without changing the positions of the cell, optics, or photodiode detector.}
\end{figure*}

Our optically detected EPR apparatus is described in detail in Ref.~\cite{Nahlawi2019}; we provide here a higher-level schematic (Fig.~\ref{fig:app}) including the important adjustments made for the present work. In brief, we used a typical SEOP setup: a circularly polarized pump laser at the $D_1$ resonance, propagating collinearly with an applied field generated by a Helmholtz pair, was used to spin-polarize the alkali-metal, which then polarized $^{129}$Xe by spin exchange. The cell temperature was maintained by a forced-air heater in an insulated aluminum oven. Temperature stability was maintained with a controller and a resistance temperature device (RTD; model F3101, Omega Engineering) placed in the oven. Quoted temperatures were based on readings of a second RTD attached to the cell surface under the same experimental conditions, which typically read $\approx 10$~$^{\circ}$C higher than the oven RTD. The second RTD was removed for actual data taking, as it tended to interfere with the lasers and add electrical noise. A frequency-tuned NMR coil was located under the cell; its only purpose was on-demand destruction of the $^{129}$Xe magnetization with a comb of pulses at the $^{129}$Xe Larmor frequency ($\approx 33$~kHz for $B_0=28$~G). The pulses were a few hundred microseconds in duration (chosen to correspond approximately to a $90^{\circ}$ flip-angle pulse) and 100-200~ms apart. Dual linearly (vertically) polarized probe lasers ($\approx 80$~mW), one for each alkali-metal, detuned $\approx 0.5$~nm from the respective $D_2$ wavelengths (780~nm for Rb and 852~nm for Cs), propagated transverse to $B_0$. A single pick-off mirror could be slid into place to replace one probe beam with the other, along identical lines through the cell, without having to adjust any of the other components. Each probe laser was mounted with an adjustable diffraction grating to form a Littrow cavity \cite{Hawthorn2001}, making the lasers tunable for several nanometers about the $D_2$ line.  The probe light emerging from the cell was passed through a linear polarizer set to 45$^{\circ}$ and a focusing lens before detection by a fast photodiode.

A two-turn, $\approx 8$~cm dia.\ EPR coil was positioned on either side of the cell and provided a weak transverse cw excitation. The coil was tuned and impedance-matched with separate component boxes exchanged manually between consecutive experiments for different alkali-metal species. The steady-state precession of the transverse magnetization modulated the Faraday rotation of the probe laser at the EPR frequency (e.g., $\approx 20$~MHz for $^{87}$Rb). The intensity of the modulation, as detected by the photodiode and homodyned with a rf mixer, was proportional to the hyperfine spectral intensity, and the spectrum of nearest-neighbor hyperfine transitions could thus be traced as a function of frequency. Using a voltage-controlled oscillator (VCO) as the rf source and appropriate feedback circuitry, we could lock the EPR frequency to the magnetic field inside the cell and thus follow the evolving nuclear magnetization. We locked to the end resonance corresponding to the helicity of pumping light and refer to this as pumping into the high-energy (HE, $\sigma +$ light) or low-energy (LE, $\sigma -$ light) Zeeman state, where Eq.~(\ref{eq:gammap}) yields the corresponding gyromagnetic ratio. We have developed a spreadsheet \cite{SM} that calculates the detailed properties of the alkali-metal hyperfine spectra as a function of applied field in the quadratic Zeeman regime, including the precise values of the gyromagnetic ratio $\gamma_{\rm A}$. We note that the EPR shifts due to $^{129}$Xe magnetization correspond to magnetic fields on the order of tens of microgauss; it was thus sufficient for our purposes to calculate $\gamma_{\rm A}$ up to terms quadratic in the applied field $B_0$ and use that single value in Eq.~(\ref{eq:ratio1}).

The basic experiment was to polarize the $^{129}$Xe repeatedly to its steady state value in the presence of polarized Rb and Cs, and to alternately observe the sudden Rb and Cs EPR frequency shifts caused by intentional rapid destruction of the $^{129}$Xe magnetization with an NMR-pulse comb. The pump laser was on continuously, and we counteracted slow drifts in the $^{129}$Xe saturation polarization and other instrumental drifts by switching back and forth between Rb and Cs EPR several times during the run. Regardless of which atom was optically pumped, the large interatomic spin-exchange cross sections assured that all alkali-metal isotopes were polarized to some degree and all three corresponding hyperfine spectra could be prominently observed across the temperature range of interest. As shown in Eq.~(\ref{eq:ratio1}), the ratio of Cs to Rb frequency shifts under these conditions is directly proportional to the corresponding $\kappa_0$ ratio that we ultimately wished to measure. A variable rf attenuator was used to reduce the rf power to the EPR coil to the point where the saturation polarization of $^{129}$Xe due to SEOP was negligibly affected by the small decrease in alkali-metal polarization compared to its value with no rf excitation.

Experiments were performed using two separate but similar apparatus. We treat these cases separately, because they each involve one of the two natural Rb isotopes, and because the respective data acquisitions were separated in time by more than 12 months; significant improvements were made in the interim, leading to a more precise measurement in the second case.

\subsection{\label{sec:case1}Case 1: $^{87}$Rb vs. $^{133}$Cs}
The $^{87}$Rb experiments were conducted at $B_0 = 28.3$~G using a 30-W diode-laser array (model M1B-795.2-50C-SS4.1, DILAS) for pumping light at the $D_1$ wavelength of 794.8~nm. The pump laser was operated at 25~A (15~W) and narrowed to 0.2-0.3~nm with a low-power tunable Littrow cavity oriented at 90~$^{\circ}$ to the main beam \cite{Chann2000}. Two similarly narrowed single diodes were used for Rb or Cs $D_2$ probe light (Thorlabs models L785P090 and L852P50, respectively). The LabVIEW (National Instruments) code for data acquisition was similar to that used in Ref.\ \cite{Nahlawi2019}: it coordinated near-simultaneous (within a few milliseconds) readings of the frequency counter (model 53220A, Agilent) and voltmeter (model 2010, Keithley) with a dwell time of $\approx 0.5$~s. The same circuitry introduced in Ref.~\cite{Nahlawi2019} was used to stabilize the Helmholtz current to a few parts in $10^5$ over the measurement time. The voltmeter measured the residual current drift across a stable $0.10$-$\Omega$ sensing resistor in series with the Helmholtz coils. The first data point in a set defined the absolute EPR frequency reference point $\nu_0$ for the measured shift. To a very good approximation, the shift $\delta \nu_i^{\prime}$ of the $i$th data point, corrected for current drift, was given by:

\begin{equation}
\label{eq:Icorr}
\Delta\nu_i^{\prime}=\nu_i-\nu_0\frac{I_i}{I_0},
\end{equation}

\noindent where $\nu_0$, $\nu_i$ and $I_0$,$I_i$ were the frequencies and currents measured at the initial and $i$th data points, respectively. Equation~(\ref{eq:Icorr}) is convenient because it does not depend on explicit knowledge of $\gamma_{\rm A}$ or the coefficient relating $B_0$ to $I$.

\begin{figure*}
\includegraphics[width=\textwidth]{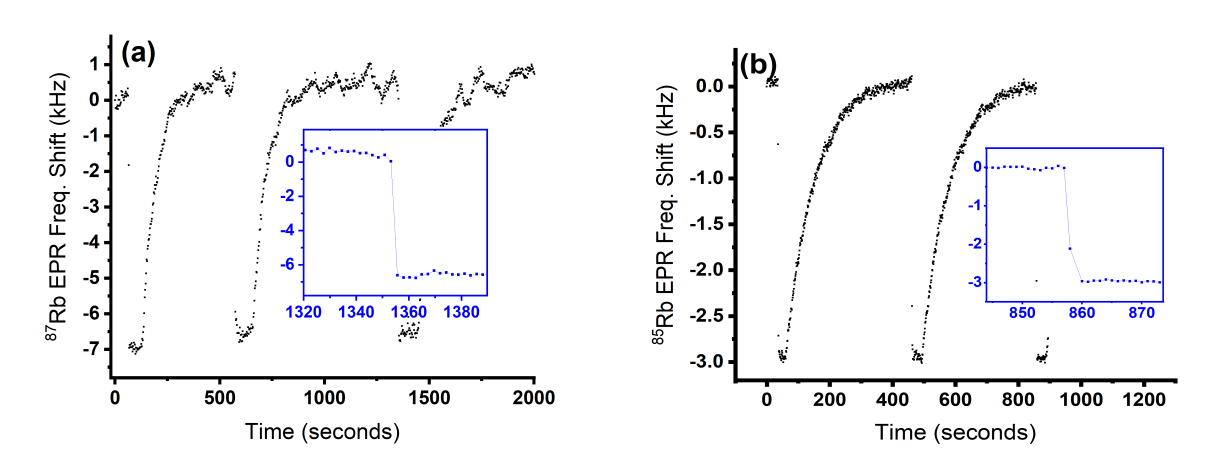}
\caption{\label{fig:cycle} Sample EPR locked-frequency-shift data for (a) $^{87}$Rb (Case 1) and (b) $^{85}$Rb (Case 2). The sharp drops in frequency shift correspond to sudden destruction of fully polarized $^{129}$Xe magnetization. The insets show the transition in more detail. In both cases, Rb is the directly pumped alkali-metal; the frequency variations characteristic of the Case-1 data are due to the light-shifts caused by fluctuating pump-laser power. These fluctuations are greatly reduced in Case 2, where we allowed the pump laser to warm up for more than a day to reach steady-state operation.}
\end{figure*}

\begin{figure}
\includegraphics[width=\columnwidth]{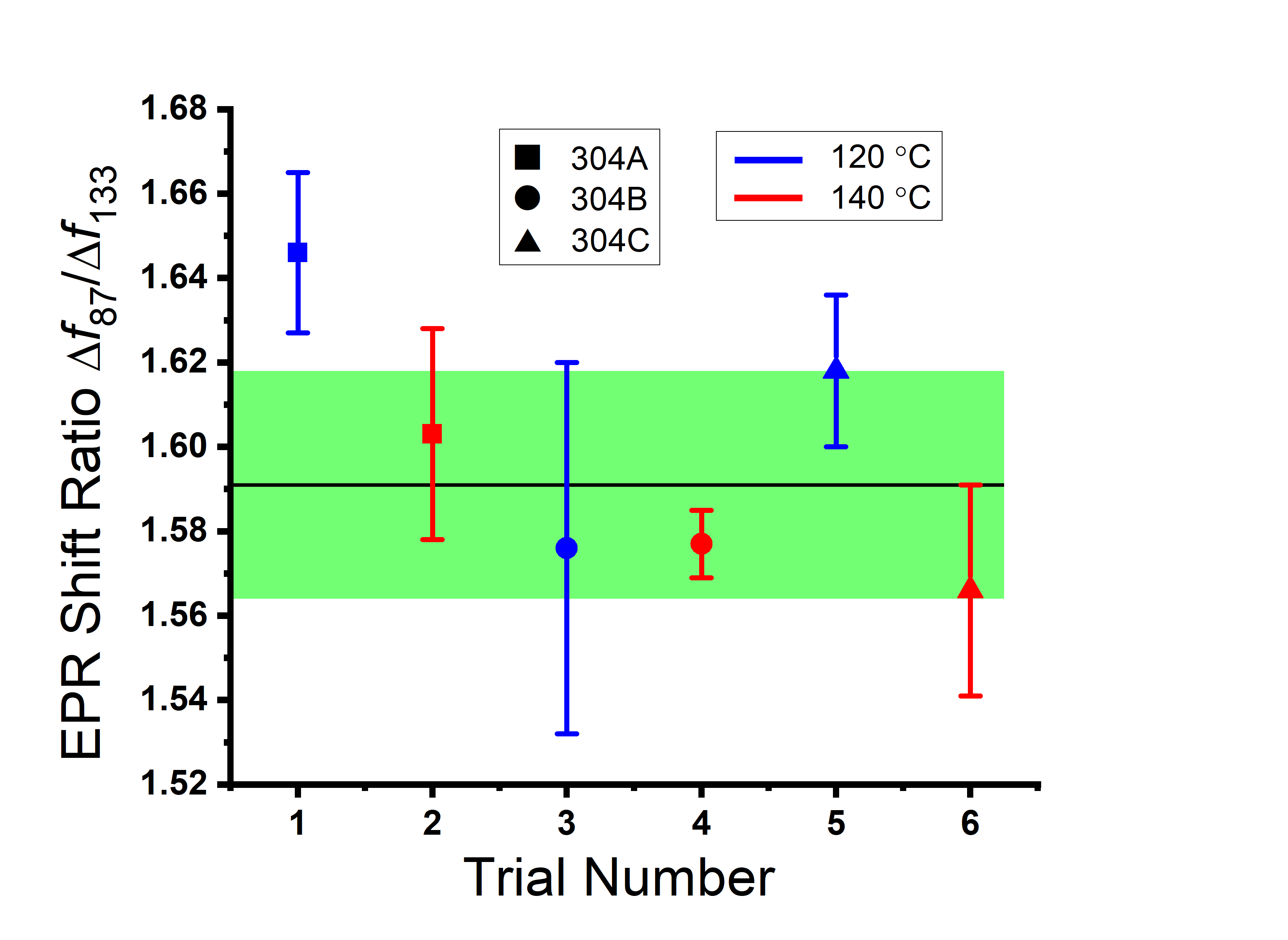}
\caption{\label{fig:87data} Plot of the ratio of $^{87}$Rb to $^{133}$Cs EPR frequency shifts resulting from the sudden destruction of the same quantity of $^{129}$Xe magnetization in three cells at two different temperatures. All data were acquired by pumping with Rb $D_1$ light at 794.8~nm into the low-energy (LE) Zeeman state. The weighted average and standard deviation are $1.591\pm 0.027$ (line and green band). The data show no obvious trends across different cells or temperatures.}
\end{figure}

Three cycles of saturation/destruction of the nuclear magnetization probed by $^{87}$Rb EPR were continuously recorded; then the probe laser was switched and three more cycles were recorded using $^{133}$Cs EPR. Example data are shown in Fig.~\ref{fig:cycle}. For analysis purposes, we recorded and averaged the ten points prior to and ten points subsequent to the NMR spin destruction, throwing out points that obviously occurred during the transition (the locking circuit can follow the sudden destruction within one or two points at most). An average shift for the three $^{87}$Rb cycles is determined and compared to the corresponding average for the three $^{133}$Cs cycles to yield a shift ratio. This entire process is repeated three times in each experimental run, yielding three independent values of the shift ratio $\Delta f_{87}/\Delta f_{133}$, from which a weighted average and standard error are calculated. Sample data sheets with embedded error calculations are included as Supplemental Material \cite{SM}. The three cells were each measured at two different temperatures to yield the six data points shown in Fig.~\ref{fig:87data}.

\begin{figure*}
\includegraphics[width=7.5in]{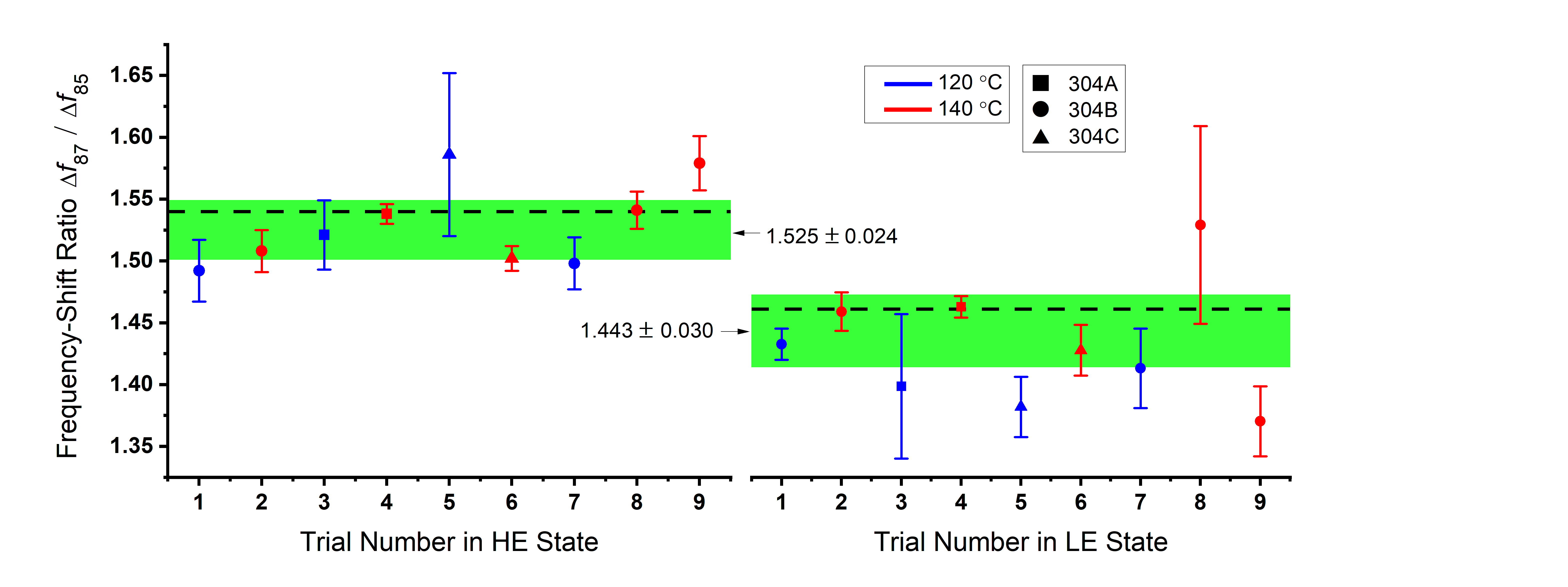}
\caption{\label{fig:8785} Plot of the ratio of $^{87}$Rb to $^{85}$Rb EPR frequency shifts for the same cells and temperatures as in Fig.~\ref{fig:87data} for both HE- and LE-state pumping. The weighted-average ratio is indicated with the uncertainty (green band) for both cases. The dashed lines (1.540 for HE and 1.461 for LE) are precisely calculated values of the expected shift ratio for each case based on the known gyromagnetic ratios (correct to linear terms in $B_0$). That we accurately measure the $\approx 5\%$ difference in the HE- and LE-state values is strong validation of the technique for determining the same ratio for the $^{87}$Rb and $^{133}$Cs EPR shifts.}
\end{figure*}

A check on our data-acquisition protocol was done by comparing the EPR shifts for $^{87}$Rb and $^{85}$Rb in exactly the same way as described above. We assume that $(\kappa_0)_{\rm RbXe}$ has negligible dependence on Rb isotope, because it depends only on the interatomic potential and the perturbed wave function of the valence electron. The difference in Rb-$^{129}$Xe reduced mass results in a $< 1\%$ difference in mean thermal collision speed; the resulting slight decrease in collision duration for $^{85}$Rb is offset by a corresponding increase in collision frequency. The shift ratio can thus be predicted precisely from $\gamma_{87}/\gamma_{85}$. At our applied field, this ratio is different by $\approx 5\%$ for the high- vs.\ low-energy end resonance. Figure~\ref{fig:8785} demonstrates our sensitivity to this small difference.


\subsection{\label{sec:case2}Case 2: $^{85}$Rb vs. $^{133}$Cs}

These experiments were conducted at $B_0 = 29.15$~G using a 35-W Rb $D_1$ (794.8~nm) VBG diode-laser array with a 35-pm line width (OptiGrate). The laser is one half of a dual-head turn-key system with wavelength and power controlled through a computer interface that actively monitors the current and junction temperature. The other half of the system is an analogous 35-W array at the Cs $D_1$ wavelength (894.4~nm); the Cs-$D_1$ laser was not used for the ratio measurments but was used to demonstrate the effects of light shifts on the hyperfine resonances (see below). Components not specified in this section are the same as for Case 1.

Improvements were made in the $B_0$-stability circuitry: thermally sensitive components were all mounted to a water-chilled aluminum slab, and the slab was covered to minimize temperature fluctuations associated with air currents. The original goal was to achieve ppm stability over the tens of minutes that it typically takes to measure SEOP transients; here, we needed stability for only about 30-60~s at a time, and these improvements practically made the current correction discussed in the previous section unnecessary \cite{fn:techrpt}. Due to our proximity to nearby experiments employing switched magnetic fields, moving elevators, and other ambient sources, we found it useful to monitor ambient field fluctuations with a fluxgate magnetometer (model FS1-100, Stefan-Mayer). The magnetometer's range is only $\pm 1$~G, but it could be positioned near a field null for the Helmholtz coil, about a coil radius ($\approx 50$~cm) away from the cell, where it was then sensitive only to ambient field fluctuations and not to changes in the Helmholtz current. Properly calibrated, field sources much further away from the cell than a coil radius could be effectively subtracted out of the EPR measurement; the LabVIEW code was updated to perform this operation in real time as frequency-shift data were being acquired \cite{fn:LVcode}.

\begin{figure*}
\includegraphics[width=\textwidth]{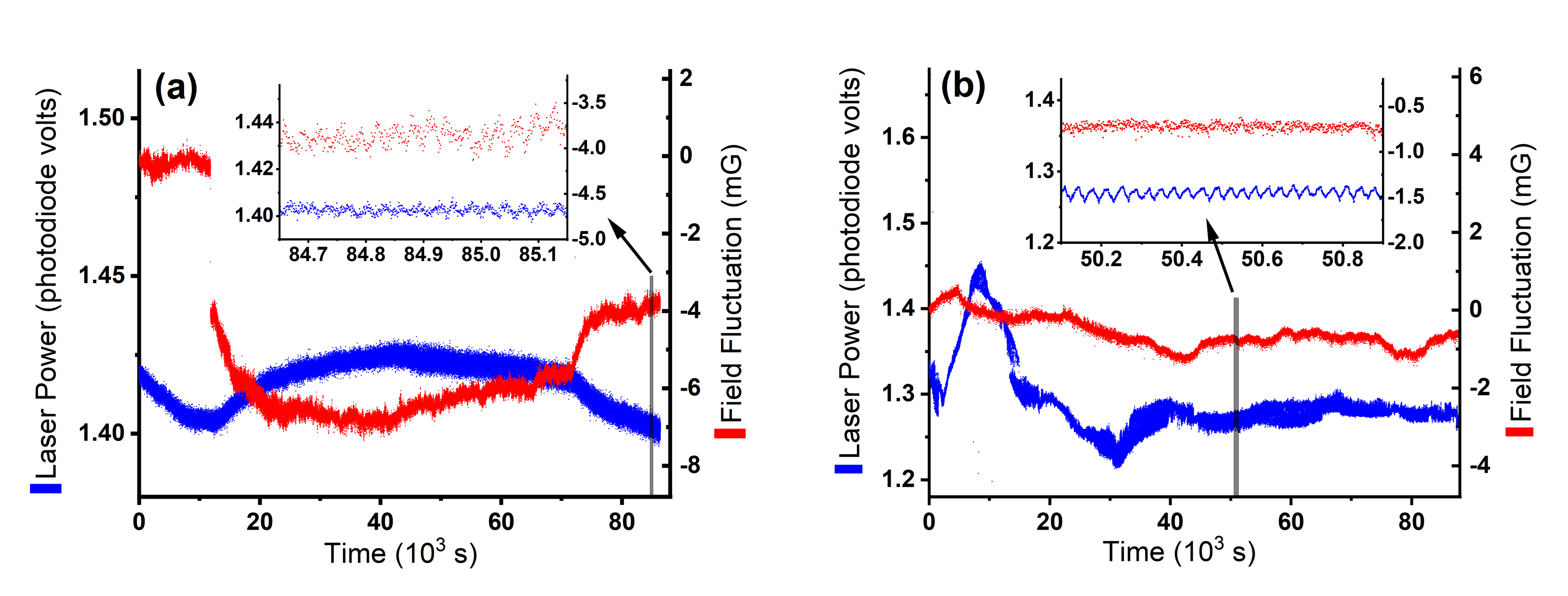}
\caption{\label{fig:PD} Plots of magnetic-field fluctuations (relative to $\approx 28.0$~G) as measured by the alkali-metal EPR frequency shift (right axis) overlayed with plots of relative pump-laser power (left axis) vs.\ time. In both (a) and (b), the pump laser operates at the Cs $D_1$ wavelength with about 28.5~W generating both Cs and Rb polarization in the hybrid vapor cell. The magnetic field is measured via the shift of the $^{133}$Cs end resonance $\ket{F,\overline{m}_F}=\ket{3,5/2}$ in (a) and the $^{87}$Rb end resonance $\ket{F,\overline{m}_F}=\ket{2,3/2}$ in (b). The cell contains no Xe or other polarizable noble-gas species.  Variations in the pump-laser power produce field fluctuations measured by EPR of the directly pumped (Cs) species (a) that are both larger than those measured by EPR of the indirectly pumped (Rb) species and correlated on both long and short (inset) time scales with the pump-laser power. In (a) the effective magnetic field associated with the light shift was opposite to the applied field $B_0$, so field fluctuations are $180^{\circ}$ out of phase with laser-power fluctuations. (The sudden jump in measured field at $\approx 11,000$~s was an anomalous glitch in the EPR acquisition.) In (b) no correlation is observed on any time scale for the indirectly pumped species.}
\end{figure*}

Improvements in signal-to-noise ratio were made after discovering that the OptiGrate laser exhibits substantial variations in output power at fixed current, particularly during its ``warm-up" phase, which may last up to 12~h. These variations led to fluctuations in the measured EPR frequency due to a variable light shift \cite{Appelt1998,Mathur1968} for whichever alkali metal was being directly pumped and polarized with $D_1$ light. We picked off a small fraction of the pumping light with a polarizing beam-splitter and measured its intensity with a photodiode; we observed a correlation between the photodiode voltage and fluctuations in the magnetic field as measured by the locked EPR frequency on time scales from tens of seconds to many hours with a cell containing no xenon (to avoid any confusion with SEOP transients or fluctuating/drifting $^{129}$Xe polarization); see Fig.\ \ref{fig:PD}a. In Fig.\ \ref{fig:PD}b, we show that there is a reduction in measured field fluctuations and no correlation with the pump-laser power if we use the locked EPR frequency of the indirectly polarized atom ($^{87}$Rb in this case). By monitoring the pump-laser power and allowing significant warm-up time, we could run the experiment during relatively stable periods to avoid significant drifts due to the light-shift effect. Similar power variations were later observed in the DILAS laser used in the $^{87}$Rb experiments (Case 1) but were not recognized or controlled for at the time. Figure~ \ref{fig:cycle} demonstrates the corresponding decrease in background noise for Case 2 compared to Case 1.

\begin{figure}
\includegraphics[width=\columnwidth]{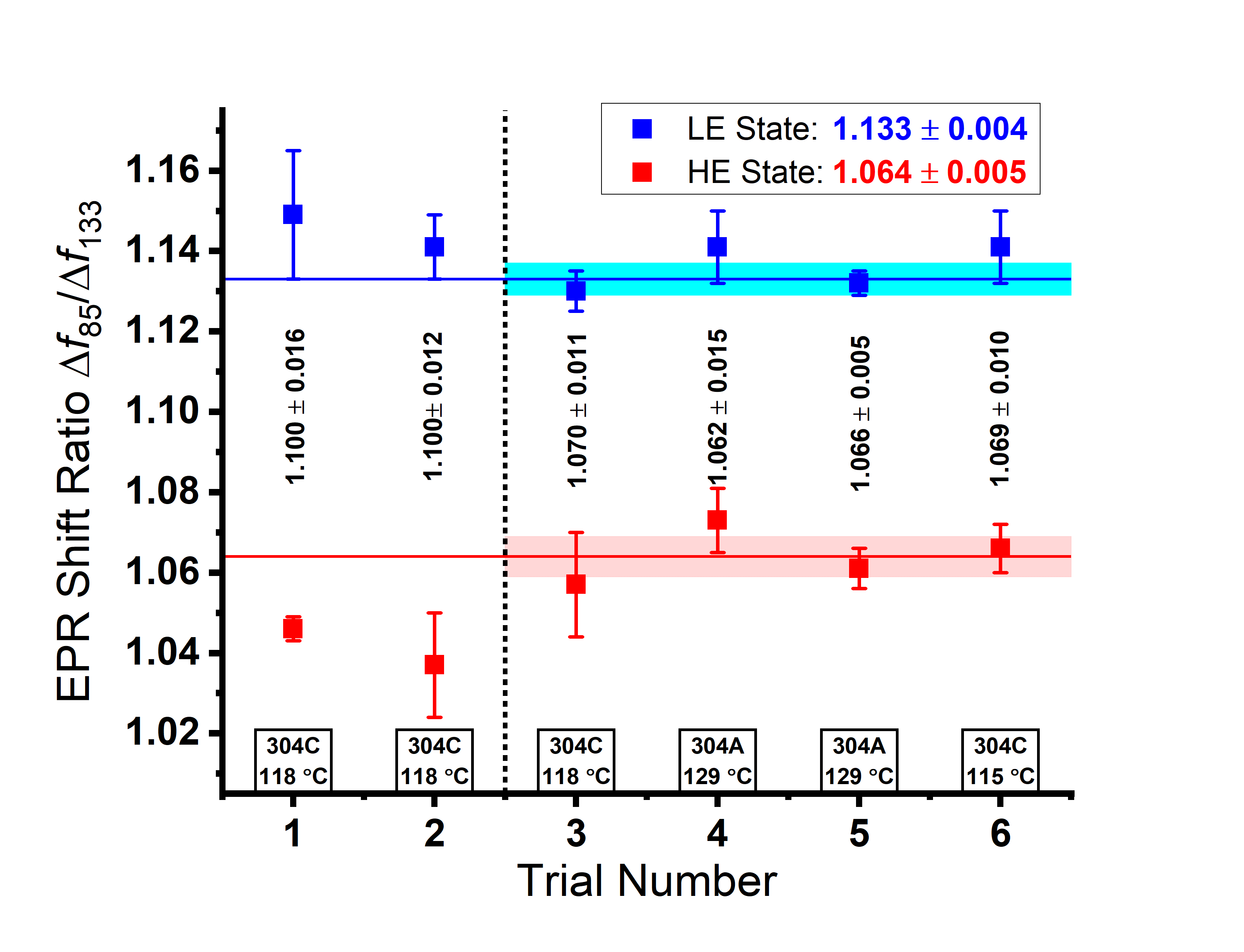}
\caption{\label{fig:85133} Plot of the ratio of EPR frequency shifts ($^{85}$Rb vs.\ $^{133}$Cs) resulting from the sudden destruction of the same $^{129}$Xe magnetization in hybrid (Rb/Cs) vapor cells. Cell designation and temperature are shown for each trial; data were acquired pumping into both the high-energy (HE) and low-energy (LE) Zeeman states ($\overline{m}_F = \pm I$). Data to the left of the vertical dashed line were acquired before we understood light-shift noise from the pump laser; only data to the right of the line are included in the analysis. The uncertainty-weighted averages with standard deviations are shown for both HE and LE pumping (lines with blue and red bands, respectively). The ratio of these values (shown in vertically between the data points for each trial) serves as a systematic check on the result of each trial, since the ratio depends only on the precisely known gyromagnetic ratios, per Eq.~(\ref{eq:ratrat}). For the results shown here, the ratio in trials 3-6 is within 1\% of the expected value of 1.060.}
\end{figure}

The data-acquisition protocol was unchanged from Case 1, except that we acquired both LE- and HE-state shift ratios in each experimental run; this allowed an immediate systematic check on the data by forming the ``ratio of ratios:"

\begin{equation}
\label{eq:ratrat}
\frac{(\Delta f_{133}/\Delta f_{85})_{\rm LE}}{(\Delta f_{133}/\Delta f_{85})_{\rm HE}}=\frac{(\gamma_{133}/\gamma_{85})_{\rm LE}}{(\gamma_{133}/\gamma_{85})_{\rm HE}},
\end{equation}

\noindent which should be the same for all data sets at a fixed $B_0$, since it depends only on the precisely known gyromagnetic ratios.  Two measurements were done on each of two different vapor cells at temperatures ranging from 115-130~$^{\circ}$C to yield the four LE and four HE data points (trials 3 through 6) shown in Fig.~\ref{fig:85133}; the plot includes data from two previous trials (1 and 2) acquired before we understood the problem with light-shift variations. The data analysis includes only the last four trials. For each of these trials, the ratio computed from the data by Eq.~(\ref{eq:ratrat}) is $<1\%$ from the expected theoretical value of 1.060 at $B_0=29.15$~G. The computed ratios are all slightly larger than but have uncertainty ranges that encompass this theoretical value.

\section{\label{sec:results}Results and Discussion}

Table~{\ref{tab2}} shows the $^{87}$Rb-EPR result (Case 1, LE state only) and the two $^{85}$Rb-EPR results (Case 2, both HE and LE states). The $\kappa_0$ ratio is computed from the weighted average of the data shown in Figs.~\ref{fig:87data} and \ref{fig:85133}, assuming that the data are independent of temperature. The weighted average of the three values in the rightmost column of Table~\ref{tab2} yields the final result:

\begin{equation}
\label{eq:ratioresult}
\frac{(\kappa_0)_{\rm CsXe}}{(\kappa_0)_{\rm RbXe}}=1.215 \pm 0.007.
\end{equation}

\begin{table}
\caption{\label{tab2}
Final results for three independent measurements of the CsXe/RbXe $\kappa_0$-ratio.}
\begin{ruledtabular}
\begin{tabular}{llllll}
&$B_0$ (G)&$\gamma_{87}/\gamma_{85}$&$\gamma_{87}/\gamma_{133}$&$\gamma_{85}/\gamma_{133}$&$\kappa_0$-ratio\\
\\[-0.9em]
\hline
\\[-0.6em]
Case 1 (LE) & 28.3 & 1.4613  &  2.0042  &  1.3715   &  $1.260 \pm 0.024$ \\
Case 2 (LE)	& 29.1 & 1.4602  &  2.0043  &  1.3726   &  $1.212 \pm 0.004$ \\
Case 2 (HE) & 29.1 & 1.5411  &  1.9948  &  1.2944   &  $1.217 \pm 0.005$ \\
\end{tabular}
\end{ruledtabular}
\end{table}

Our previous measurement of $(\kappa_0)_{\rm RbXe}=518\pm 8$ \cite{Nahlawi2019} combined with the present result allows a determination of the Cs-$^{129}$Xe enhancement factor: $(\kappa_0)_{\rm CsXe}= 629 \pm 10$. Both of these results rely on the previous precise measurement of the Rb-$^3$He enhancement factor made by Romalis and co-workers \cite{Romalis1998}. Both are about 25\% smaller than the values theoretically predicted by Walker \cite{Walker1989}; however, our independent measurement of the ratio of these two parameters is in excellent agreement with the ratio determined from Ref.\ \cite{Walker1989}.

A previous measurement of $(\kappa_0)_{\rm CsXe}$ based on $^{129}$Xe NMR frequency shifts made by Fang and co-workers \cite{Fang2014} yielded $702 \pm 41$ at 80~$^{\circ}$C and $653 \pm 20$ at 90~$^{\circ}$C. Given the uncertainties in these measurements, they are not inconsistent with our result or with the absence of temperature dependence observed in our work. Temperature dependence for the heavier noble gases is expected to be weak at best, since $\kappa_0$ depends more on the location of the steep repulsive core of the interatomic potential than on the well depth \cite{Walker1989}.

With respect to the precision of our results, we have done a conventional statistical analysis assuming that the results of all measurements are normally distributed. The sample data sheet included in Supplemental Material \cite{SM} shows the exact formulas used to process weighted means and expected uncertainties. Our data show no trend with temperature, and the measured shift ratio would be relatively insensitive to temperature, since any weak dependence is almost certainly in the same direction for both Rb and Cs EPR shifts. Various systematic effects, many which we did not understand or control for until the measurements of the $^{87}$Rb-Cs $\kappa_0$-ratio in Case 2 were undertaken, were present intermittently and to different degrees throughout the Case 1 measurements. We have no reason to believe that these effects consistently skewed the data one way or the other; they simply caused the background noise in reading the EPR frequency shift to be larger or smaller depending on time of day, whether elevators were running, whether a neighboring lab switched a magnet on or off, etc. We thus take the trials as independent, with the uncertainty determined by the background noise for each trial, but given these varying experimental conditions and the relatively few number of data points, we use the more conservative weighted standard deviation as our quoted uncertainty for each of the three results shown in Table~{\ref{tab2}}. The final result is mostly determined by the Case 2 measurements, where experimental conditions were better understood and controlled.

Because we are using nominally spherical cells and $\kappa_0$ is at least several hundred, the effects of a finite through-space field generated by the $^{129}$Xe magnetization due to cell asphericity and the small stem (``pull-off") created when the cell is flame-sealed away from the glass manifold are negligible \cite{Nahlawi2019}. The main source of systematic error is the variable light-shift discussed in Sec.\ \ref{sec:case2}. Light-shift variations were uncontrolled in Case 1, degrading the precision of that result, although we have no reason to suspect that the accuracy was affected. We thus include the Case 1 result in a straightforward weighted average along with the more precise results from Case 2. Allowing the OptiGrate laser to warm up for more than 12~h worked well enough for these short measurement times to improve precision significantly in Case 2. In future work on spin-exchange rate coefficients, we will describe a ``locking in the dark" technique that enables stable frequency-locking measurements over longer time scales that are impervious to light-shift effects.

Knowledge of the alkali-metal densities in our cells was not necessary for these experiments, and we did not directly measure them. Given the volume ratios of Rb to Cs metal (Table~\ref{tab1}), we can use vapor pressure curves and Raoult's Law to estimate densities of both metals to be in the range $5\times 10^{12}$ to $5\times 10^{13}$ over the temperature range studied \cite{SM}. Using known values of the Rb-Cs spin-exchange cross section \cite{Gibbs1967}, we estimate the Rb-Cs spin-exchange rate $\Gamma_{\rm SE}$ in our cells ranged from $4\times 10^{4}$ to $4\times 10^{5}$~s$^{-1}$. We also measured the alkali-metal spin-destruction rates $\Gamma_{\rm A}$ in cells with similar gas composition and total pressure to be much slower---on the order of $10^3$~s$^{-1}$. The condition $\Gamma_{\rm SE}\gg\Gamma_{\rm A}$ should mean that all the hyperfine spectra are observable with a single probe laser near the $D_2$ resonance of either alkali-metal atom. This was pointed out to us by E. Babcock after we had conducted all of our experiments with two separate probe lasers. We subsequently tried to observe the indirectly probed Cs (Rb) EPR hyperfine spectrum with the Rb (Cs) probe laser but were unsuccessful.  In similar work with Rb-K hybrid cells and $^3$He (no Xe), Babcock, et al.\ \cite{Babcock2003,BabcockThesis} worked with higher alkali-metal densities and much lower spin-destruction rates, better satisfying the above condition and perhaps leading to easier observation of the indirectly probed EPR spectrum. A single probe laser would have made for a simpler and more efficient experiment, and we are uncertain as to why it didn't work, but we remain confident that our approach using two separate probe lasers produced reliable reproducible data.

\begin{acknowledgments}
The authors thank glassblower A. Babino for vapor cell and glass manifold fabrication and E. Babcock for helpful discussions; we acknowledge the U.S. National Science Foundation grant PHY-1708048 for support of this work.
\end{acknowledgments}


\begin{thebibliography}{99}

\bibitem{Walker1997} T. G. Walker and W. Happer, Rev. Mod. Phys. \textbf{69}, 629 (1997).

\bibitem{Gentile2017} T. R. Gentile, P. J. Nacher, B. Saam, and T. G. Walker, Rev. Mod. Phys. \textbf{89}, 045004 (2017).

\bibitem{Meersman2014XeBook} T. Meersmann and E. Brunner, eds., \textit{Hyperpolarized Xenon-129 Magnetic Resonance: Concepts, Production, Techniques and Applications} (Royal Society of Chemistry, Cambridge, 2015).

\bibitem{Stupic2011} K.F. Stupic, Z.I. Cleveland, G E. Pavlovskaya, and T. Meersmann, J. Magn. Reson. \textbf{208}, 58 (2011).

\bibitem{Pavlovskaya2005} G.E. Pavlovskaya, Z.I. Cleveland, K.F. Stupic, R.J. Basaraba, and T. Meersmann, Proc. Natl. Acad. Sci. USA, \textbf{102}, 18275 (2005).

\bibitem{Allmendinger2019} F. Allmendinger, I. Engin, W. Heil, S. Karpuk, H.-J. Krause, B. Niederl\"{a}nder, A. Offenh\"{a}usser, M. Repetto, U. Schmidt, and S. Zimmer, Phys. Rev. A \textbf{100}, 022505 (2019).

\bibitem{Sachdeva2019} N. Sachdeva, I. Fan, E. Babcock, M. Burghoff, T.E. Chupp, S. Degenkolb, P. Fierlinger, S. Haude, E. Kraegeloh, W. Kilian, S. Knappe-Gr\"{u}neberg, F. Kuchler, T. Liu, M. Marino, J. Meinel, K. Rolfs, Z. Salhi, A. Schnabel, J.T. Singh, S. Stuiber, W.A. Terrano, L. Trahms, and J. Voigt, Phys. Rev. Lett. \textbf{123}, 143003 (2019).

\bibitem{Almasi2018} J. Lee, A. Almasi, and M. Romalis, Phys. Rev. Lett. \textbf{120}, 161801 (2018).

\bibitem{Bulatowicz2013} M. Bulatowicz, R. Griffith, M. Larsen, J. Mirijanian, C.B. Fu, E. Smith, W.M. Snow, H. Yan, and T.G. Walker, Phys. Rev. Lett. \textbf{111}, 102001 (2013).

\bibitem{Molway2021} M.J. Molway, L. Bales-Shaffer, K. Ranta, D. Basler, M. Murphy, B.E. Kidd, A.T. Gafar, J. Porter, K. Albin, B.M. Goodson, E.Y. Chekmenev, M.S. Rosen, W.M. Snow, J. Ball, E. Sparling, M. Prince, D. Cocking, M.J. Barlow, arXiv:2105.03076 [physics.atom-ph].

\bibitem{Tardiff2014} E.R. Tardiff, E.T. Rand, G.C. Ball, T.E. Chupp, A.B. Garnsworthy, P. Garrett, M.E. Hayden, C.A. Kierans, W. Lorenzon, M.R. Pearson, C. Schaub, and C.E. Svensson,  E.R. Tardiff, E.T. Rand, and G.C. Ball, Hyperfine Interact. \textbf{225}, 197 (2014).

\bibitem{Kitano1988} M. Kitano, F.P.Calaprice, M.L.Pitt, J. Clayhold, W. Happer, M. Kadar-Kallen, M. Musolf, G.Ulm, K. Wendt, T. Chupp, J.Bonn, R.Neugart, E.Otten, and H.T.Duong. Phys. Rev. Lett. \textbf{60}, 2133 (1988).

\bibitem{Driehuys1996} B. Driehuys, G.D. Cates, E. Miron, K. Sauer, D.K. Walter, and W. Happer, Appl. Phys. Lett. {\textbf 69}, 1668 (1996).

\bibitem{Ruset2006} I.C. Ruset, S. Ketel, F.W. Hersman, Phys. Rev. Lett. {\bf 96}, 053002 (2006).

\bibitem{Nikolau2013} P.\ Nikolaou, A.M.\ Coffey, L.L.\ Walkup, B.M.\ Gust, N.\ Whiting, H.\ Newton, S.\ Barcus, I.\ Muradyan, M.\ Dabaghyan, G.D.\ Moroz, M.S.\ Rosen, S.\ Patz, M.J.\ Barlow, E.Y.\ Chekmenev, and B.M.\ Goodson, Proc. Natl. Acad. Sci. USA {\bf 110}, 14150 (2013).

\bibitem{Roos2015} J.E. Roos, H.P. McAdams, S.S. Kaushik, and B. Driehuys, Magn. Reson. Imaging Clin. N. Am. \textbf{23}, 217 (2015).

\bibitem{Ruppert2019} K. Ruppert, K. Qing, J.T. Patrie, T.A. Altes, and J.P. Mugler 3rd, Acad Radiol. \textbf{26}, 355 (2019).

\bibitem{Grist2021} J.T. Grist, M. Chen, G.J. Collier, B. Raman, G. AbuEid, A. McIntyre, V. Matthews, E. Fraser, L.-P. Ho, J.M. Wild, F. Gleeson, Radiology (published online) https://doi.org/10.1148/radiol.2021210033

\bibitem{Whiting2011} N. Whiting, N.A. Eschmann, B.M. Goodson, and M.J. Barlow, Phys. Rev. A \textbf{83}, 053428 (2011).

\bibitem{Franzen1959} W. Franzen, Phys. Rev.\ {\bf 115}, 850 (1959).

\bibitem{Nelson2001} I.A. Nelson and T.G. Walker, Phys. Rev. A {\bf 65}, 012712 (2001).

\bibitem{Appelt1999} S. Appelt, A.B. Baranga, A.R. Young, and W. Happer,
Phys. Rev. A {\bf 59}, 2078-84 (1999).

\bibitem{Appelt1998} S. Appelt, A.B. Baranga, C.J. Erickson, M.V. Romalis, A.R. Young, and W. Happer, Phys. Rev. A {\bf 58}, 1412 (1998).

\bibitem{Korver2015} A. Korver, D. Thrasher, M. Bulatowicz, and T.G. Walker, Phys. Rev. Lett.\ {\bf 115}, 253001 (2015).

\bibitem{Ma2011} Z.L. Ma, E.G. Sorte, and B. Saam, Phys. Rev. Lett. {\bf 106}, 193005 (2011).

\bibitem{Nahlawi2019} A.I. Nahlawi, Z.L. Ma, M.S. Conradi, and B. Saam, Phys. Rev. A \textbf{100}, 053415 (2019).

\bibitem{Romalis1998} M.V. Romalis and G.D. Cates, Phys. Rev. A \textbf{58}, 3004 (1998).

\bibitem{Schaefer1989} S.R. Schaefer, G.D. Cates, T.-R. Chien, D. Gonatas, W. Happer, and T.G. Walker, Phys. Rev. A {\bf 39}, 5613 (1989).

\bibitem{Breit1931} G. Breit and I. I. Rabi, Phys. Rev. \textbf{38}, 2082 (1931).

\bibitem{SteckRb87-2015} D.A. Steck, ``Rubidium 87 D Line Data," available online at http://steck.us/alkalidata (revision 2.1.5, 13 January 2015).

\bibitem{SteckRb85-2013} D.A. Steck, ``Rubidium 85 D Line Data," available online at http://steck.us/alkalidata (revision 2.1.6,
20 September 2013).

\bibitem{SteckCs133-2010} D.A. Steck, ``Cesium D Line Data," available online at http://steck.us/alkalidata (revision 2.1.4, 23
December 2010).

\bibitem{Jacob2002} R.E. Jacob, S.W. Morgan, and B. Saam, J. Appl. Phys. {\bf 92}, 1588 (2002).

\bibitem{Killian1926} T.J. Killian, Phys. Rev.\ {\bf 27}, 578 (1926).

\bibitem{CRC} David R. Lide (ed), CRC Handbook of Chemistry and Physics, 84th Edition, online version. CRC Press. Boca Raton, Florida, 2003; Section 4, Properties of the Elements and Inorganic Compounds; Vapor Pressure of the Metallic Elements

\bibitem{Raoult1886} F.-M. Raoult, Comptes Rendus Acad. Sci.\ {\bf 104}, 1430-1433 (1886).

\bibitem{SM} See Supplemental Material at http://link.aps.org/ supplemental/XX.XXXX/PhysRevLett.XXX.xxxxxx.

\bibitem{Hawthorn2001} C.J. Hawthorn, K.P. Weber, and R.E. Scholten, Rev. Sci. Instrum. {\bf 72}, 4477 (2001).

\bibitem{Chann2000} B. Chann, I. Nelson, and T. G. Walker, Opt. Lett. \textbf{25}, 1352 (2000).

\bibitem{fn:techrpt} We intend to publish a technical report with details of this design.

\bibitem{fn:LVcode} The latest version of our LabVIEW code and documentation is available on request from the corresponding author.

\bibitem{Mathur1968} B.S. Mathur, H. Tang, and W. Happer, Phys. Rev., \textbf{171}, 11 (1968).

\bibitem{Walker1989} T.G. Walker, Phys. Rev. A {\bf 40}, 4959 (1989).

\bibitem{Fang2014} J. Fang, S. Wan, and Y. Chen,
Chin. Phys. B {\bf 23}, 063401 (2014).

\bibitem{Gibbs1967} H.M. Gibbs and R.J. Hull, Phys. Rev. \textbf{153}, 132 (1967).

\bibitem{Babcock2003} E. Babcock, I. Nelson, S. Kadlecek, B. Driehuys, L.W. Anderson, F.W. Hersman, and T.G. Walker, Phys. Rev. Lett. \textbf{91}, 123003 (2003).

\bibitem{BabcockThesis} E. Babcock, Ph.D. Thesis, University of Wisconsin, 2005.

\end{thebibliography}
\end{document}